\documentclass[iop]{emulateapj}
\usepackage{amsmath}
\usepackage{footnote}
\usepackage{natbib}
\usepackage[percent]{overpic}

\bibpunct{(}{)}{;}{a}{}{,}
\usepackage{hyperref}
\hypersetup{
    colorlinks,
    citecolor=blue,
    filecolor=blue,
    linkcolor=blue,
    urlcolor=blue
}
\slugcomment{Resubmitted to The Astrophysical Journal after incorporating comments from the referee.}
\shortauthors{Allen et al.}

\begin{document}

\title{Spectral Softening Between Outburst and Quiescence In The Neutron Star Low-Mass X-Ray Binary SAX J1750.8-2900}
\author{Jessamyn L. Allen\altaffilmark{1,2}}
\author{Manuel Linares\altaffilmark{3} \altaffilmark{4}}
\author{Jeroen Homan\altaffilmark{2}}
\author{Deepto Chakrabarty\altaffilmark{1,2}}

\altaffiltext{1}{Department of Physics, Massachusetts Institute of Technology, 77 Massachusetts Avenue, Cambridge, MA 02139, USA; allenjl@mit.edu}
\altaffiltext{2}{Kavli Institute for Astrophysics \& Space Research, MIT, 70 Vassar St., Cambridge, MA 02139, USA}
\altaffiltext{3}{Instituto de Astrof{\'i}sica de Canarias, c/ V{\'i}a L{\'a}ctea s/n, E-38205 La Laguna, Tenerife, Spain}
\altaffiltext{4}{Universidad de La Laguna, Dept. Astrof{\'i}sica, E-38206 La Laguna, Tenerife, Spain}

\begin{abstract}
Tracking the spectral evolution of transiently accreting neutron stars between outburst and quiescence probes relatively poorly understood accretion regimes. Such studies are challenging because they require frequent monitoring of sources with luminosities below the thresholds of current all-sky X-ray monitors. We present the analysis of over 30 observations of the neutron star low-mass X-ray binary SAX J1750.8-2900 taken across four years with the X-ray telescope aboard \emph{Swift}. We find spectral softening  with decreasing luminosity both on long ($\sim$1 year) and short ($\sim$days to week) timescales. As the luminosity decreases from $4\times10^{36}$ erg s$^{-1}$ to $ \sim1\times10^{35} $ erg s$^{-1}$ (0.5-10 keV), the power law photon index increases from from 1.4 to 2.9. Although not statistically required, our spectral fits allow an additional soft component that displays a decreasing temperature as the luminosity decreases from $4 \times 10^{36} $ to $6 \times 10^{34}$ erg s$^{-1}$. Spectral softening exhibited by SAX J1750.8-2900 is consistent both with accretion emission whose spectral shape steepens with decreasing luminosity and also with being dominated by a changing soft component, possibly associated with accretion onto the neutron star surface, as the luminosity declines.
\end{abstract}

\small

\section{Introduction}
\label{sec:intro}
Low-mass X-ray binaries (LMXBs) are powerful and strongly variable X-ray sources that provide a means to study accretion flows over a wide luminosity range and on relatively short timescales. Neutron star LMXBs consist of a neutron star accreting via Roche-lobe overflow from its low-mass ($\lesssim 1 M_{\odot}$) companion. In transient LMXBs, the accretion rate onto the neutron star surface varies by many orders of magnitude, but the systems spend the majority of their time in a low-luminosity, $10^{31}-10^{33}$ erg s$^{-1}$ (0.5-10 keV unabsorbed luminosity, denoted by $L_X$), state of quiescence during which there is little or no accretion. In outburst, these systems undergo intense accretion for periods of weeks to months (to years in the extreme case of quasi-persistent transients), with typical X-ray luminosities of $10^{36}-10^{38}$ erg s$^{-1}$. In this work, we focus on the intermediate luminosity range between outburst and quiescence of $10^{34}-10^{36}$ erg s$^{-1}$.

Limited instrumental sensitivity and/or observation time makes it difficult to study neutron star transients during their, often rapid (lasting less than several weeks), transition from outburst to quiescence. Hence, the accretion processes and emission sources in this luminosity regime are not well-understood. However, as more and more neutron star binaries are being monitored between outburst and quiescence, there is growing evidence that these systems may share the common behavior of spectral softening with decreasing luminosity. Spectral softening between outburst and quiesence has been documented in detail for transient black hole binaries \citep{yuanbh,wugubh,plotkin} and for individual neutron star transients \citep{padillaxtej1719,igrj17494,m28soft}. It has recently been proposed as a feature exhibited by most neutron star transients between luminosities of $10^{36}$ and $10^{34}$ erg s$^{-1}$ \citep{wijnandssoft}. In black hole systems, the softening is commonly attributed to radiatively inefficient accretion flows. The underlying mechanism(s) responsible for the softening in neutron star systems is likely more complicated due to the presence of the neutron star surface and magnetic field.

A better understanding of accretion at luminosities between $10^{34}$ and $10^{36}$ erg s$^{-1}$ may provide insight into accretion processes at even lower luminosities while the systems are in quiescence. In addition to thermal emission from the neutron star surface, many quiescent transients display a non-thermal, high-energy spectral tail above 2 keV which can be modeled with a power law ($\Gamma$ = 1-2)\footnote{where $\Gamma$ is defined as $dN/dE \propto E^{-\Gamma}$}. Nonthermal and thermal emission may be variable both in their total flux and relative flux contribution over short and long timescales, as has been found in the well-studied neutron star transient Cen X-4 \citep{cenx4quiesc}. Several quiescent transients have also exhibited flares, sudden increases in flux reaching ten or more times the typical quiescent emission levels and exponentially decaying over the course of a few days \citep{fridriksson_2011,degenaarks1741}. The physical origin of the non-thermal emission in quiescence is not entirely understood, but is generally attributed to ongoing accretion \citep{campana1998}, while flares are interpreted as sporadic increases in the accretion rate onto the neutron star, although the physics of disk instabilities at low accretion rates is not well-understood.

More frequent and deeper observations of neutron star LMXBs between their quiescent and outburst states, as well as observations of flares in quiescence, are critical to constraining the physical origins of the hard and soft spectral contributions, together with their evolution in time. Such work is necessary to understand the nature of accretion at all luminosities. In particular, it can help in identifying the accretion regimes that heat the neutron star crust, which has important implications for equilibrium temperature measurements of the crust and studies of the thermal evolution of transient neutron stars.

SAX J1750.8-2900 was first detected in outburst in 1997 by the Wide Field Cameras on board \emph{BeppoSAX} and has since been detected in outburst an additional 3 times with both long (4-5 months) and short ($\le$ 1 month) outburst durations. The presence of type I X-ray bursts revealed that the source is a neutron star LMXB \citep{natalucci}. Using \emph{RXTE} observations of its 2001 outburst, the second recorded for this source, \citet{galloway} analyzed photospheric radius expansion bursts to establish a distance of $6.79 \pm 0.14$ kpc for hydrogen-poor burning. In March 2008, SAX J1750.8-2900 entered its third known outburst \citep{atel1425} with a reported flux of 230 mCrab corresponding to a luminosity $3.1 \times 10^{37}$ erg s$^{-1}$ (2-10 keV, d = 6.79 kpc), similar to the peak flux exhibited in previous outbursts. By August 2008 the source fell below the \emph{RXTE} detection threshold and was reported to be returning to quiescence \citep{atel1662}. \emph{SuperAGILE} reported a 60 seconds long burst on October 9, 2008 \citep{atel1775} indicating the system had rebrightened. The return to outburst was also seen in two \emph{Swift} XRT observations on October 1$^{st}$ and 8$^{th}$ \citep{atel1777}. By mid-February 2009, SAX J1750.8-2900 had returned to quiescence \citep{lowell}.

Over a year after the end of its last reported outburst, SAX J1750.8-2900 was observed for the first time in quiescence in April 2010 with \emph{XMM}with a bolometric luminosity of $1.05 \pm 0.12 \times 10^{34} \ (D/6.79\ $ kpc)$^2$ erg s$^{-1}$ ($L_X = 8.9 \times 10^{33}$ erg s$^{-1}$), making it the most luminous quiescent neutron star known \citep{lowell}. The spectrum was well-fit with a neutron star atmosphere model with a high effective temperature ($kT_{\mathrm{eff}}^{\infty} = 148$ eV). Under the assumption the source had undergone no additional heating (i.e. accretion) since outburst, \citet{lowell} claimed that the neutron star crust and core were in equilibrium and, thus, the high surface temperature reflected a hot neutron star core.

In the two years after the \emph{XMM} observation, SAX J1750.8-2900 exhibited a short outburst ($<$ 1 month) and a flare in quiescence. A February 2011 increase in activity reported by \emph{IBIS/INTEGRAL} \citep{atel3170} was confirmed as a faint outburst by \emph{Swift} XRT with $L_X = 1.05 \times 10^{36}$ erg s$^{-1}$ for a source distance of 6.79 kpc \citep{atel3181}. In March 2012, the detection by \emph{Swift} XRT of a low-level flare (with characteristic $L_X = (3-4) \times 10^{34}$ erg s$^{-1}$ assuming d = 6.79 kpc) lasting between 5 and 16 days indicated the system may undergo small accretion events between outbursts \citep{saxjflare}.

No orbital parameters are known for this neutron star binary and no observations of the companion star exist except for reports of a near-IR candidate counterpart  \citep{atel1472}. The line-of-sight towards SAX J1750.8-2900 suffers from high-extinction ($A_V$\footnote{http://irsa.ipac.caltech.edu/applications/DUST/} = 13.0-15.1), consistent with the high column density, $N_H > 10^{22}$, which may limit observations of the optical counterpart.

In this work we present a detailed spectral analysis of the neutron star LMXB SAX J1750.8-2900, observed with \emph{Swift} XRT over the course of four years. We have tracked the X-ray spectral evolution of SAX J1750.8-2900 over three orders of magnitude in luminosity between outburst and quiescence, detecting spectral softening with decreasing luminosity. We interpret the spectral softening as being generated by changes in soft and hard accretion-powered emission as the mass accretion rate decreases, and we investigate the potential crustal heating effects due to accretion at sub-outburst rates.

\section{Observations \& Analysis}
\begin{figure}
\begin{center}
\includegraphics[scale = 0.75]{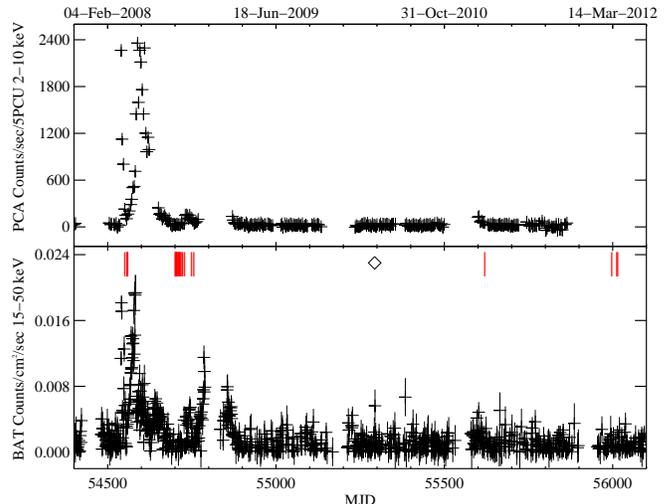}
\caption{Light curves for SAX J1750.8-2900 from \emph{RXTE} PCA (top) and the \emph{Swift} BAT monitor (bottom). In April 2010, indicated by the diamond symbol, SAX J1750.8-2900 was observed in quiescence with \emph{XMM} and \citet{lowell} reported a quiescent luminosity of $L_X = 9 \times 10^{33}$ erg s$^{-1}$. The red vertical line symbols in the bottom panel indicate all 27 \emph{Swift} XRT observations with source detections, analyzed herein. Within this set of observations, SAX J1750.8-2900 had luminosities between $10^{34}$ and $10^{37}$ erg s$^{-1}$ (see Table \ref{tab:groupfits}).  \protect{\label{fig:rxteswiftlc}} }
\end{center}
\vspace{-2mm}
\end{figure}

We analyzed all \emph{Swift} XRT \citep{swiftxrt} observations where SAX J1750.8-2900 was in the field of view (FOV), which yielded 27 source detections out of a total of 38 observations. Five observations had SAX J1750.8-2900 more than 7 arcminutes off-axis (i.e., inside but near the edge of the FOV), but we included them in our analysis as luminosity upper limits. The data set spans over four years of activity, between March 2008 and September 2012, with a total exposure time of almost 70 ks. The long-term soft and hard X-ray light curves from SAX J1750.8-2900 measured with \emph{RXTE} PCA and \emph{Swift} BAT \citep{batcurve}, respectively, covering the timespan of our \emph{Swift} observations are shown in Figure \ref{fig:rxteswiftlc}.  The majority of our \emph{Swift} XRT observations cover the tail end of the 2008 outburst and the return to outburst just several weeks later in October; the remaining observations primarily occurred in February and March 2012. A log of the \emph{Swift} XRT observations is presented in Table \ref{tab:obsummary}.

%Light curves from RXTE & Swift

\subsection{ Source Detection \& Spectral Extraction}
\label{sec:detections-spec-extraction} 

All observations were re-processed with the \texttt{xrtpipeline} tool (v.0.12.6). The source was identified in photon counting (PC) mode observations with the \emph{sosta} tool set to a signal-to-noise threshold limit of 2.0 in \emph{XImage} (v.4.5.1).

Pile-up corrections were made for PC mode observations with count rates greater than $0.5$ s$^{-1}$ within a circular region with a 20-pixel (47 arcsec) radius centered on the source. The extent of the piled-up region was determined following the \emph{Swift} XRT Pile-Up thread\footnote{http://www.swift.ac.uk/analysis/xrt/pileup.php
}. In these observations, an annulus with an inner radius set to exclude the piled-up pixels (radius of 3 pixels for mild pile-up and up to 12 pixels for the most severe case of pile-up, ObsID 31166001) and an outer radius of 30 pixels (45 pixels for 31166001) was used for the source region. The background spectra were extracted from two circular regions, both with radii of 45 pixels. All PC mode observations with count rates between 0.004 and 0.5 s$^{-1}$ (i.e., not affected by pile-up) had spectra extracted from a circular region with a 20-pixel radius centered on the source, and the background was extracted from two circular regions, both with radii of 35 pixels. Potentially nearby  X-ray sources (within 15 arcseconds) were identified using \emph{Simbad}\footnote{http://simbad.u-strasbg.fr/simbad/} and were avoided in any of the source and background regions.

Window timing (WT) mode observations only incur pile-up for count rates above 100 s$^{-1}$; all SAX J1750.8-2900 count rates were below 10 s$^{-1}$. Thus, no pile-up corrections were required in this mode. Both the source and background spectra in WT mode were extracted from circular regions with 30-pixel radii.

%Observation Details Table
\begin{centering}
\begin{table}
\centering 
\caption{Exposures and background corrected count rates for all observations with and without SAX J1750.8-2900 detections. 95\% upper count rate limits (indicated by the '$<$' symbol) are given in cases where the source was not detected above the SNR threshold ( $\leq 2$). The '*' indicates the far off-axis observations ($>$ 7') with SAX J1750.8-2900 within the FOV. \label{tab:obsummary} }
\tabcolsep=0.11cm
\tabletypesize{\footnotesize}
\begin{tabular}{ccccc}
\hline
\hline
ObsID & Observation & Mode & Exposure & Count Rate \\
 	& Date		 & 	& (ks) 	& (s$^{-1}$) \\
\hline
31166001 & 2008-03-17 & PC &  1.0 &  $ 3.36 \pm  0.06 $ \\
31174001 & 2008-03-24 & PC &  0.7 &  $ 1.56 \pm  0.05 $ \\
31174001 & 2008-03-24 & WT &  0.3 &  $ 5.71 \pm  0.14 $ \\
31174002 & 2008-03-25 & PC &  0.9 &  $ 1.38 \pm  0.04 $ \\
31174002 & 2008-03-25 & WT &  0.1 &  $ 3.96 \pm  0.23 $ \\
31174003 & 2008-08-14 & PC &  1.9 &  $ 0.06 \pm  0.01 $ \\
31174004 & 2008-08-15 & PC &  1.9 &  $ 0.04 \pm  0.01 $ \\
31174005 & 2008-08-16 & PC &  2.0 &  $ 0.06 \pm  0.01 $ \\
31174006 & 2008-08-17 & PC &  2.1 &  $ 0.05 \pm  0.01 $ \\
31174007 & 2008-08-18 & PC &  2.1 &  $ 0.10 \pm  0.01 $ \\
31174008 & 2008-08-19 & PC &  1.7 &  $ 0.12 \pm  0.01 $ \\
31174009 & 2008-08-20 & PC &  2.0 &  $ 0.08 \pm  0.01 $ \\
31174010 & 2008-08-24 & PC &  1.8 &  $ 0.06 \pm  0.01 $ \\
31174011 & 2008-08-25 & PC &  2.1 &  $ 0.13 \pm  0.01 $ \\
31174012 & 2008-08-26 & PC &  2.5 &  $ 0.12 \pm  0.01 $ \\
31174013 & 2008-08-27 & PC &  1.9 &  $ 0.11 \pm  0.01 $ \\
31174014 & 2008-08-28 & PC &  1.7 &  $ 0.08 \pm  0.01 $ \\
31174015 & 2008-08-29 & PC &  5.9 &  $ 0.050 \pm  0.0003 $ \\
31174016 & 2008-09-03 & PC &  2.3 &  $ 0.14 \pm  0.01 $ \\
31174017 & 2008-09-05 & PC &  0.7 &  $ 0.05 \pm  0.01 $ \\
31174018 & 2008-09-11 & PC &  1.8 &  $ 0.07 \pm  0.01 $ \\
31174019 & 2008-10-01 & PC &  2.0 &  $ 0.74 \pm  0.02 $ \\
31174020 & 2008-10-08 & PC &  1.9 &  $ 0.62 \pm  0.02 $ \\
31174021 & 2011-02-19 & PC &  3.9 &  $ 0.46 \pm  0.01 $ \\
31174023 & 2011-09-30 & PC & 0.5 & $<$0.018 \\
31174024 & 2012-02-14 & PC & 3.8 & $<$0.004 \\
31174025 & 2012-02-26 & PC & 2.6 & $<$0.003 \\
31174026 & 2012-02-29 & PC & 0.3 & $<$0.023 \\
31174027 & 2012-03-03 & PC & 3.2 & $0.002 \pm 0.001 $ \\
31174028 & 2012-03-06 & PC & 2.8 & $<$0.003 \\
31174030 & 2012-03-17 & PC & 3.1 & $0.018 \pm 0.003 $\\
31174031 & 2012-03-20 & PC & 1.0 & $0.006 \pm 0.003 $ \\
31174032 & 2012-03-22 & PC & 1.0 & $<$0.006 \\
43649001 & 2012-05-22 & PC & 0.5 & $<$0.019* \\
32427001 & 2012-06-17 & PC & 0.8 & $<$0.009* \\
32427002 & 2012-06-22 & PC & 0.7 & $<$0.008* \\
32427003 & 2012-08-15 & PC & 0.1 & $<$0.042* \\
32427004 & 2012-09-04 & PC & 0.5 & $<$0.010* \\
Quiescence & - & PC & 12.3 &  $0.0011 \pm  0.0004$ \\
Flare & - & PC &  5.1 &  $ 0.012 \pm  0.002 $ \\
\hline
\end{tabular}
\end{table}
\end{centering}

After extracting the source and background spectra in \emph{XSelect} (v.2.4b) with standard event grades 0-2 for WT and 0-12 for PC modes, we created exposure maps and ancillary files using the \texttt{xrtexpomap} and \texttt{xrtmkarf} tools, respectively, using the most recent RMFs (v.14 for PC mode and v.15 for WT mode) and enabling the vignetting and PSF correction options. Whenever possible, we grouped energy bins to a minimum of 20 counts per bin using the \texttt{grppha} tool. In cases where there were only dozens of counts, we binned as low as 5 counts per bin. We confirmed that the spectral parameters obtained with binning to fewer than 20 counts per bin were consistent with the spectral parameters from fits of unbinned spectra.

\subsubsection{Non-detections and combined observations}
\label{sec:secnondetect}
In 11 out of a total of 38 observations, we did not detect our source. We were able to place upper limits on SAX J1750.8-2900's activity using a circular extraction region with a radius of 15 pixels centered on the published \emph{Chandra} coordinates for SAX J1750.8-2900 \citep{atel1490}. Following the prescription of \citet{gehrels}, we placed the 95\% upper limits on the source count rate (denoted by $<$ in Table \ref{tab:obsummary}).

We combined observations with no source detection and observations with faint source detections at low intensities (with count rates $<0.02$ s$^{-1}$) during the February and March 2012 period to increase the source counts and perform spectral fits. Observations 31174030, 31174031 and 31174032, corresponding to March 17-22, exhibited elevated count rates compared to previously exhibited quiescent levels ($\lesssim 0.002$ s$^{-1}$); we combined these three observations to form a 'Flare' spectrum. We formed a 'Quiescent' spectrum by summing all observations immediately prior to the flare with count rates (including upper limits) less than $0.004$ s$^{-1}$ for a total exposure time of 12.3 ks (corresponding to ObsIDs 31174024-25 and 31174027-28).  Spectra for the flare and quiescence were generated by summing observations' event files in \emph{XSelect} followed by summing their exposure maps in \emph{XImage}. The source and background spectra were extracted following the same procedure previously outlined for PC mode observations with clear source detections using a circle with a 10-pixel radius as the source region.

\begin{figure}
\begin{center}
\begin{overpic}[width=0.5\textwidth]{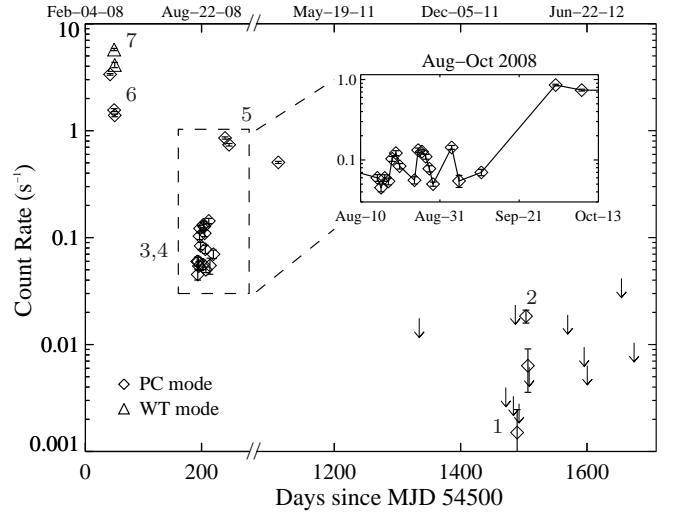}
 \put (18,68) {7}  \put (18,60) {6}  \put (20,37) {3,4} \put (77,30) {2} \put (35,57) {5} \put (72,11) {1}   \end{overpic}
\vspace{-1mm}
\caption{SAX J1750.8-2900's count rate light curve: The downward arrows indicate the upper 95\% count rate limits for all observations without source detections (see Section \ref{sec:secnondetect}), and the inset shows the activity between August and October 2008 where the bulk of our observations occur. The numbers indicate the approximate groups of observations with similar spectral properties and luminosities. \protect{\label{fig:ctrt}} }
\end{center}
\end{figure}

The source intensity as a function of the observation date for our entire data set is shown in Figure \ref{fig:ctrt}.
 
%Count rate plot

\subsection{Spectral Analysis}
\label{sec:specanalysis}

All spectral analysis was performed using \emph{XSpec} v.12.7.0 \citep{xspecref}. We used the \texttt{tbabs} multiplicative model to account for interstellar absorption and set our abundances to those of \citet{wilmsabund} and used the cross-sections from \citet{verner}. All quoted luminosities are the 0.5-10 keV unabsorbed luminosities (unless a different energy range is specified), calculated with the \texttt{cflux} convolution model assuming a source distance of $6.79$ kpc \citep{galloway}. The Eddington fraction ($L_X / L_{Edd}$) assumes an Eddington luminosity, L$_{Edd}$, of  $2 \times 10^{38}$ erg s$^{-1}$ for a 1.4 M$_{\odot}$ neutron star with a hydrogen-rich photosphere. Errors on fit parameters correspond to the 90\% confidence bounds ($\Delta \chi^2 = 2.71$) and all chi-squares, $\chi^2 \ (dof)$, are reduced.

\subsubsection{Hardness Ratio}
\label{sec:hratio}

In order to study the evolution of the average spectral slope or "hardness" in a model-independent way, we calculated the hardness ratio (HR) for each observation. We have defined the hardness ratio as the ratio of hard (2-10 keV) to soft counts (0.5-2 keV), which we calculated within \emph{XSpec} from background-corrected source spectra binned to 1 count per bin to avoid having empty bins. 

\emph{Swift} XRT analysis guides\footnote{http://www.swift.ac.uk/analysis/xrt/digest\_cal.php} warn of soft spectral 'bump' below 1 keV for highly absorbed sources ($N_H >  10^{22}$ cm$^{-2}$) when using the standard WT mode event grades 0-2 selection. We investigated the grades 0-2 versus grade 0 extractions for our two WT mode observations and found no significant difference in the hardness ratios between the two grade selections (ObsID 31174001 HR =  $5.7 \pm 0.4$ vs $5.95 \pm 0.5$, 31174002 HR = $6.6 \pm 1.3$ vs $6.1 \pm 1.1$ for grades 0 and 0-2, respectively) or in parameters obtained from spectral fits (for example, a simple absorbed power law). Thus, we used the standard WT event grades 0-2 extraction in our analysis.

\subsubsection{Column Density Constraints and Spectral Modeling}
\label{sec:specmodel}

The column density towards SAX J1750.8-2900 is poorly constrained; published values range from $N_{H} = (2.5 - 6.0) \times 10^{22}$ cm$^{-2}$, henceforth noted as $N_{H, 22} = 2.5 - 6.0$ \citep{natalucci,lowell} and are strongly dependent on the spectral model used. From a quiescent \emph{XMM} observation, \citet{lowell} obtained $N_{H,22} = 5.9 \pm 0.5$ with the \texttt{NSATMOS} model \citep{nsatmos}, while a blackbody fit yielded $N_{H,22} = 4.0^{+1.1}_{-0.9}$. 

Fitting individual observations with an absorbed power law with no fixed inputs led to large uncertainties in the spectral parameters. In order to obtain better constraints, we fit groups of observations with a single model and tied certain parameters between all observations within the group. We formed the groups by initially fitting all individual observations with an absorbed power law fixing the column density to $N_{H,22} = 5$ (the average of the values obtained by \citealt{lowell}) while leaving the photon index and normalization free to vary; we then sorted the observations by their 2-10 keV unabsorbed luminosities\footnote{We first attempted to sort the observations by their 0.5-10 keV unabsorbed luminosity, but found a weaker correlation between the luminosity and spectral shape (i.e. photon index). For highly absorbed sources ($N_{H,22} \gtrsim 3$) and soft spectra ($\Gamma \gtrsim 2$), a power law model leads to large 0.5-2 keV unabsorbed fluxes due to the divergence of the power law flux at low energies. Thus, we found a stronger spectral dependence on luminosity when we excluded the low-energy flux contribution by looking at the 2-10 keV unabsorbed luminosity.}. Plotting the 2-10 keV unabsorbed luminosities against the powerlaw photon index, we found 7 distinct clusters of observations in the photon index-luminosity parameter space and divided our set of observations into 7 spectral and luminosity groups, shown in Table \ref{tab:groupfits}.

We also attempted a single component fit with a blackbody model (\texttt{bbodyrad}), but found it did not provide acceptable fits across our dataset's entire luminosity range. At high luminosities ($L_X >10^{37}$ erg s$^{-1}$, the blackbody yielded a significantly worse fit than a power law ($\chi^2=1.2$ vs 0.9). Additionally, column densities required for the blackbody fits  ($N_{H,22} \lesssim 2.5$) were lower than even the low-end of published estimates.

We performed a new power law fit  for each group, tying the photon index while leaving the normalization free. For the two brightest groups (6 and 7) we left the absorption parameter free (but tied if there were multiple observations in the group) and found consistent values for the column density, $N_{H, 22} = 4.2 \pm 0.4$ and  $4.4 \pm0.4$. We fixed all future column densities to $N_{H, 22} = 4.3$, the average value from our two brightest groups.

%BB  - FEB/MARCH2012
\begin{figure}
\begin{center}
\includegraphics[scale = 0.75]{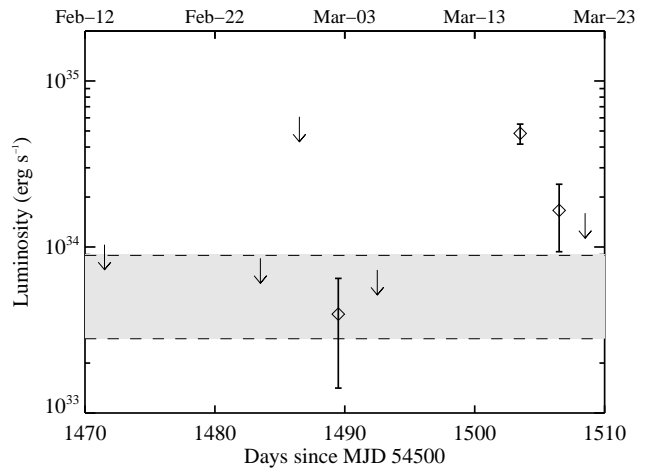}
\caption{The 0.5-10 keV unabsorbed luminosity estimates for  SAX J1750.8-2900 between February and March 2012, including the flare reported by \citet{saxjflare} which begins as early as March 6. Both luminosity upper limits (indicated by the downward pointing arrows) and luminosity estimates for observations with source detections (diamond symbols) were calculated from the count rate-flux conversion based on a blackbody with $kT_{\mathrm{eff}}^{\infty}$ = 331 eV. The shaded region is bounded by the quiescent luminosities reported by \citet{lowell} for blackbody ($L_X=2.8\times10^{33}$ erg s$^{-1}$ and neutron star atmosphere models ($L_X=8.9\times10^{33}$ erg s$^{-1}$). \protect{ \label{fig:posinglefitplot2012}} }
\end{center}
\vspace{-2mm}
\end{figure}

While the group fits provided tight constraints on SAX J1750.8-2900's average spectral changes with luminosity, to examine the source's variability on shorter timescales we fit the individual observations with an absorbed power law, fixing the column density while leaving the photon index and normalization free to vary. Due to insufficient source counts, we were unable perform spectral fits for ObsIDs 31174024-31174032 in an attempt to track the changes in the photon index before, during and after the flare. Instead, we used a count rate-flux conversion to estimate the system's luminosity. Using \emph{WebPimms} (v4.6a)\footnote{http://heasarc.gsfc.nasa.gov/Tools/w3pimms.html}, we converted the observed count rate/upper limit to a luminosity based on the published blackbody quiescent fit  result of \citet{lowell} ($kT_{\mathrm{eff}}^{\infty} = 331$ eV and $N_{H,22} = 4$), shown in Figure \ref{fig:posinglefitplot2012}. The quiescent and flare spectra formed by combining observations  (Section \ref{sec:secnondetect}), however, had enough counts to perform power law fits.

While the power law model worked well for both the group and individual fits (we obtained $\chi^2 \sim 1$), we investigated whether multiple spectral components were present by performing a multicomponent model fit to the grouped observations where our statistics were good enough to constrain additional parameters. We added a blackbody component (\texttt{bbodyrad}) to the existing powerlaw. When tying the blackbody temperature and normalization across all observations within a group, we were able to constrain the soft component's temperature, emitting area (i.e. the blackbody model's normalization) and its fractional contribution to the total unabsorbed flux for groups 3-6.

%Hardness Ratio Plot
\begin{figure}
\begin{center}
\includegraphics[scale = 0.75]{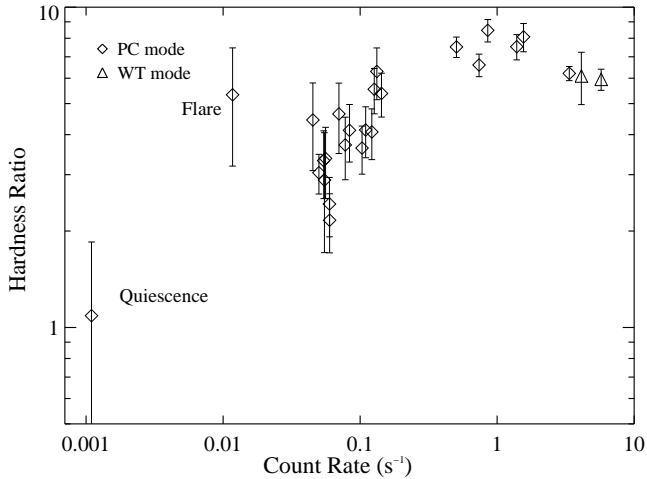} 
\caption{Hardness ratio as a function of count rate: As the count rate decreases between $1.5-0.05$ s$^{-1}$, the hardness ratio also decreases, indicating the source softens as its intensity declines. This behavior is intrinsic to the source as the hardness ratio is independent of a spectral model. The source may soften further towards quiescence (down to a count rate of ~ $0.001$ s$^{-1}$) but we discuss the limitations of interpreting the softening with low count spectra in Section \ref{sec:powerlawresults}. \protect{\label{fig:hratio}} }
\end{center}
\vspace{-3mm}
\end{figure}

\section{Results}
In our set of observations, SAX J1750.8-2900's count rate and luminosity span over 3 orders of magnitude with ranges of 0.001  to 6 s$^{-1}$ and $L_X$ between $10^{34}$ to $10^{37}$ erg s$^{-1}$. Observed at the beginning of the March 2008 outburst with $L_X = (4 - 30) \times 10^{36}$ erg s$^{-1}$, the source declined in intensity and luminosity in August 2008 ($L_X \sim 1 \times 10^{35}$ erg s$^{-1}$) until it re-brightened in October 2008. In March 2012, SAX J1750.8-2900's activity was below the \emph{RXTE} PCA and \emph{Swift} BAT sensitivity thresholds, but its quiescent activity ($\lesssim 10^{34}$ erg s$^{-1}$) was still marginally detectable with \emph{Swift} XRT. Our observations were all well-fit by a pure power law, and in some cases, we were able to fit groups of observations with a power law plus blackbody model.

\subsection{Spectral Softening, Hardening and Variability}
\label{sec:softening}

We find a clear softening towards lower luminosities of the X-ray spectrum of SAX J1750.8-2900. Hardness ratios and photon indices decrease and increase, respectively, as the system's unabsorbed luminosity decreases.

In Figure \ref{fig:hratio}, as the source's intensity decreases, $1.5 \rightarrow 0.05$ counts s$^{-1}$, the spectrum softens as the hardness ratio decreases by nearly a factor of 3, from $8 \rightarrow 3$. The results presented in Figure \ref{fig:hratio} use hardness ratios and count rates for pile-up corrected spectra and are therefore independent of the spectral model used. Further softening at even lower count rates ($<0.05$ s$^{-1}$) is discussed in the next section.

\subsubsection{Model 1: Power law behavior}
\label{sec:powerlawresults}

\begin{figure}
\begin{center}
\begin{overpic}[width=0.475\textwidth]{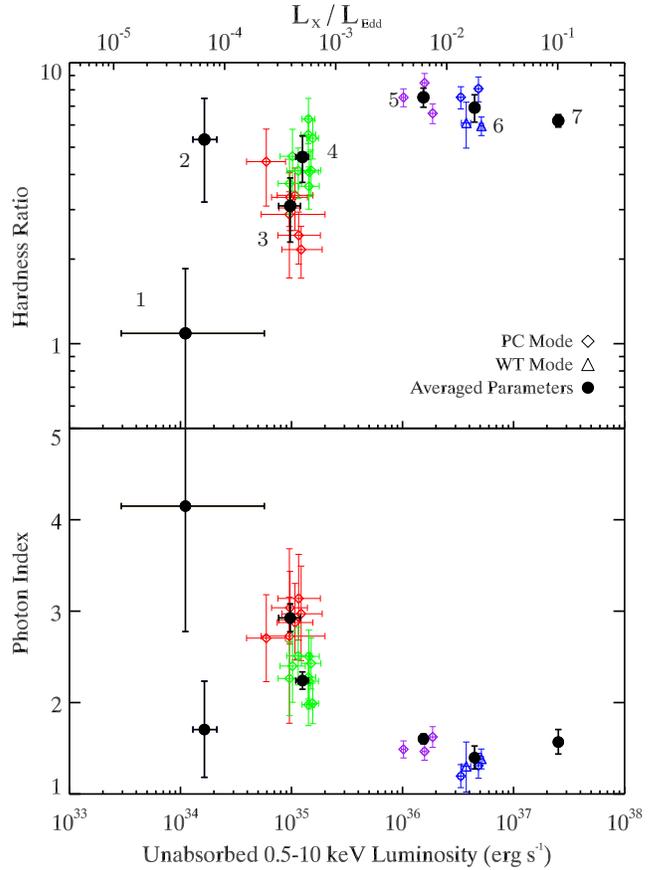}
 \put (15,65) {1}  \put (20,81) {2}  \put (29,72) {3} \put (37,82) {4} \put (44,88) {5} \put (56,85) {6}  \put (65,86) {7}  \end{overpic}
\vspace{-2mm}
\caption{\textbf{Model 1, Power Law:} Hardness ratios (model-independent) and photon indices from a power law fit versus the unabsorbed 0.5-10 keV luminosity, also given in terms of the Eddington fraction ($L_X/L_{Edd}$). Empty symbols indicate values for the individual observations and the symbol color represents its luminosity group. Black filled circles are the results for the fits of the grouped observations (with the group numbers adjacent). Spectral softening, first evident in the hardness ratio-count rate relation in Figure \ref{fig:hratio}, is further supported by the anti-correlation and correlation of the hardness ratio and photon index, respectively, with the luminosity between $4 \times 10^{36}$ and $10^{35}$ erg s$^{-1}$.  \protect{\label{fig:pobinnedplot}} }
\end{center}
\vspace{-1mm}
\end{figure}

\begin{figure}
\begin{center}
\includegraphics[scale = 0.75]{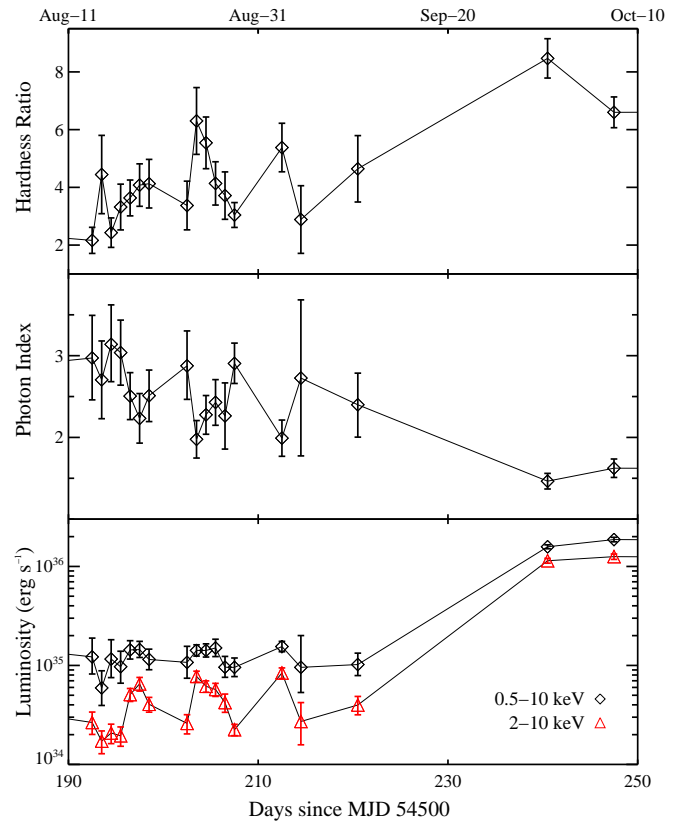}
\caption{\textbf{Model 1, Power Law:} Between August and October 2008, the hardness ratios and photon indices obtained with a power law model show signs of correlation and anti-correlation, respectively, with the 0.5-10 keV unabsorbed luminosity, indicating the spectrum softens (hardens) as the luminosity decreases (increases) on timescales from days-weeks. Also plotted in the bottom panel is the 2-10 keV unabsorbed luminosity, which shows stronger variation in time.\protect{ \label{fig:posinglefitplot2008}} }
\end{center}
\vspace{-2mm}
\end{figure}

The softening behavior is also evident in the power law spectral parameters both when observations are fit as groups and individually. Power law parameters obtained from group fits are shown in Table \ref{tab:groupfits} and plotted in Figure \ref{fig:pobinnedplot}, while power law fits to individual spectra are contained in Table \ref{tab:posinglefits} and plotted in Figures \ref{fig:pobinnedplot} and \ref{fig:posinglefitplot2008} for the full data set and the August to October 2008 period, respectively. As seen in Figure \ref{fig:pobinnedplot}, as the 0.5-10 keV unabsorbed luminosity decreases, $L_X = 4 \times 10^{36} \rightarrow 1 \times 10^{35}$ erg s$^{-1}$, the photon index increases from 1.4 to 2.9.

%Binned Power Law & PL + BB Fits Parameters- Table
\begin{table*}[ht!]
\centering
\caption{Power Law and Power Law + Blackbody Fits - Grouped Observations \label{tab:groupfits}}
\tiny
\tabcolsep=0.07cm
\begin{tabular}{c c c | c c c c | c c c c r l c l c | c}
\tableline
\noalign{\smallskip}
\tableline
 & & & \multicolumn{3}{c}{Model 1:\ \texttt{tbabs * powerlaw} } &  & & & \multicolumn{5}{c}{Model 2:\ \texttt{tbabs * (powerlaw + bbodyrad)}} & & & \\
Group & Low \tablenotemark{a} & High \tablenotemark{a} & N$_{H,22}$ & $\Gamma$ & L$_X$ & $\chi^2$ (dof) & N$_{H,22}$ & $\Gamma$ & kT \tablenotemark{c} & BB-Norm& L$_X$ & $F_{PL}$\tablenotemark{d}  & $F_{BB}$ \tablenotemark{d}  & $\chi^2$ (dof) & F-Test \tablenotemark{e}  & HR  \\
\tableline
1 & 0.001 & 0.01 &  4.3 & $ 4.1 \pm  1.6$ & $ 0.11^{+ 0.46}_{- 0.08}$ &  15.1 (18) \tablenotemark{b} & &\multicolumn{7}{c}{--}  &  & 1.1 $\pm$  0.8 \\
2 & 0.01 & 0.1 &  4.3 & $ 1.7 \pm  0.5$ & $ 0.16^{+ 0.05}_{- 0.03}$ &  68.4 (59)  \tablenotemark{b} & &\multicolumn{7}{c}{--}  & &   5.3 $\pm$  2.1 \\
3 & 0.1 & 0.3 &  4.3 & $ 2.9 \pm  0.2$ &  $ 0.97^{+ 0.23}_{- 0.21}$&  0.9 (108) &  4.3 & $ 2.5^{+ 0.8}_{- 1.2}$ & $ 0.49^{+ 0.15}_{- 0.07}$ & $ 6.2^{+ 8.1}_{- 4.9}$ & $ 0.6^{+ 0.2}_{- 0.1}$ & $ 66 \pm  32$ \% & $ 34 \pm  32$ \%  &  0.9 (106) & 2e$^{-3}$  & 3.1 $\pm$  0.8 \\
4 & 0.3 & 5 &  4.3 & $ 2.2 \pm  0.1$ &  $ 1.26^{+ 0.17}_{- 0.16}$ &  1.0 (144) &  4.3 & $ 1.6^{+ 0.4}_{- 0.4}$ & $ 0.55^{+ 0.06}_{- 0.05}$ & $ 6.7^{+ 3.5}_{- 3.1}$ & $ 1.0^{+ 0.1}_{- 0.1}$ & $ 62 \pm  12$ \% & $ 38 \pm  12$ \%  &  0.8 (142) & 7e$^{-9}$ &  4.6 $\pm$  0.9 \\
5 & 5 & 20 &  4.3 & $ 1.6 \pm  0.1$ &  $ 15.42^{+ 0.66}_{- 0.66}$ &  1.0 (231) &  4.3 & $ 1.6^{+ 0.2}_{- 0.2}$ & $ 1.04^{+ 0.17}_{- 0.15}$ & $ 4.4^{+ 3.4}_{- 1.9}$ & $ 14.4^{+ 0.7}_{- 0.7}$ & $ 79 \pm  7$ \% & $ 21 \pm  7$ \%  &  0.8 (229) &   5e$^{-2}$ &7.5 $\pm$  0.6 \\
6 & 20 & 100 & $ 4.4\pm 0.4$ & $ 1.4 \pm  0.1$ &  $ 44.57^{+ 3.67}_{- 3.34}$ &  1.0 (216) &  4.3 & $ 1.6^{+ 0.3}_{- 0.3}$ & $ 1.89^{+ 0.74}_{- 0.52}$ & $ 1.7^{+ 1.8}_{- 1.5}$ & $ 44.6^{+ 3.5}_{- 3.0}$ & $ 77 \pm  11$ \% & $ 23 \pm  11$ \%  &  1.0 (215) & 8e$^{-6}$ &  6.9 $\pm$  0.8 \\
7 & 100 & 500 & $ 4.2\pm 0.4$ & $ 1.6 \pm  0.1$ &  $253.13^{+ 19.75}_{- 16.00}$ &  0.9 (139) & &\multicolumn{7}{c}{--}  & &   6.2 $\pm$  0.3 \\
\tableline
\end{tabular}
\footnotetext[1]{Group luminosity bounds in units $10^{35}$ erg s$^{-1}$ for the 2-10 keV unabsorbed luminosity obtained with a power law fit.}
\footnotetext[2]{C-statistic}
\footnotetext[3]{Blackbody temperature (keV)}
\footnotetext[4]{Percent contributions of the power law and blackbody components to the total 0.5-10 keV unabsorbed luminosity.}
\footnotetext[5]{F-test probability that the added component (i.e. blackbody) improves the fit by chance.}
\end{table*}

Below an intensity of 0.05 s$^{-1}$ and a luminosity of $\sim 5 \times 10^{34}$ erg s$^{-1}$, the spectral parameters and hardness ratios suffer from large uncertainties due to the low number of counts, making the softening trend less obvious. \citet{lowell} and \citet{saxjflare} found that SAX J1750.8-2900's 2010 and 2012 quiescent spectra, respectively, were well-fit with a thermal spectral model (either \texttt{NSATMOS} or \texttt{bbodyrad}). In particular, for the  deep \emph{XMM} quiescent observation where more stringent limits on the nonthermal component could be placed, no additional power law was necessary and a power law component had a maximum contribution of 4\% to the total flux \citep{lowell}. The thermal nature of the quiescent spectrum combined with our low hardness ratio (HR = 1.1) suggest it is possible that the softening continues towards quiescence near $L_X \sim 10^{34}$ erg s$^{-1}$. The flare, however, appears inconsistent with the softening behavior. It exhibits a higher hardness ratio and lower photon index than the trend would predict given the event's luminosity, but the error bars are large for both parameters, so we are unable to claim that the flare is an exception to the softening trend with such limited statistics 

%Binned Power Law Fit Parameters - Plot

In addition to the general trend of softening as the luminosity decreases over an order of magnitude, we found softening occurred on very short timescales (on the order of days to weeks), as well as spectral \emph{hardening} as the luminosity increased on similar timescales. While the 2-10 keV luminosity was more strongly variable than the 0.5-10 keV luminosity (see Figure \ref{fig:posinglefitplot2008}), between August and October 2008, which corresponds to groups 3 and 4, hardness ratios and photon indices were correlated and anti-correlated, respectively, with the 0.5-10 keV luminosity\footnote{As mentioned in \ref{sec:specmodel}, the suppressed variation in the 0.5-10 keV unabsorbed luminosity may be due to the large 0.5-2 keV unabsorbed flux contribution in a power law fit to highly absorbed, soft spectra.} which varied between $5.9 \times 10^{34}$ and $1.6 \times 10^{35}$ erg s$^{-1}$.

For example, SAX J1750.8-2900 softened between August 25-29 (ObsIDs 31174011-4015); as the intensity, hardness ratio ($6.3 \rightarrow 3$), and 0.5-10 keV luminosity decreased ($L_X = (1.42 \rightarrow 0.96) \times 10^{35}$ erg s$^{-1}$) while the photon index increased ($2.0 \rightarrow 2.9$). Immediately after the softening, the spectrum hardened; between August 29 - September 3 (ObsIDs 31174015-4016), SAX J1750.8-2900's luminosity increased to $1.6 \times 10^{35}$ erg s$^{-1}$, while the hardness ratio increased to 5.4 and the photon index decreased to $\Gamma = 2.0$.

%Simpl(BB) Fit - AUG/SEPT 2008

%Power Law Single Fit Table
\begin{table}[!ht]
\centering
\caption{Model 1: Power Law Fits - Individual Observations \label{tab:posinglefits} }
\scriptsize
\tabcolsep=0.03cm
\begin{tabular}{c c | c c c c c}
\tableline
\noalign{\smallskip}
\tableline
	&	&	&	\multicolumn{2}{c}{\texttt{tbabs * powerlaw} }   &  & \\
Group & ObsID & N$_{H,22}$ & $\Gamma$ & L$_X$ \tablenotemark{a} & $\chi^2$ (dof) & HR  \\
\tableline
1 & Quiescence &  4.3 &  $ 4.1 \pm  1.6$ &  $ 0.11^{+ 0.46}_{- 0.08}$ &  15.1 (18) \tablenotemark{b} &  1.1$\pm$  0.8 \\
\hline
2 & Flare &  4.3 & $ 1.7 \pm  0.5$ &  $ 0.16^{+ 0.05}_{- 0.03}$ &  68.4 (59) \tablenotemark{b} &  5.3 $\pm$  2.1 \\
\hline
3 & 31174004 &  4.3 & $ 2.7 \pm  0.5$ &  $ 0.59^{+ 0.29}_{- 0.20}$ &  1.3 (15) &  4.4 $\pm$  1.4 \\
 & 31174005 &  4.3 & $ 3.1 \pm  0.5$ &  $ 1.16^{+ 0.66}_{- 0.41}$ &  0.8 (21) &  2.4 $\pm$  0.5 \\
 & 31174015 &  4.3 & $ 2.9 \pm  0.2$ &  $ 0.96^{+ 0.23}_{- 0.19}$ &  1.5 (13) &  3.0 $\pm$  0.4 \\
 & 31174006 &  4.3 & $ 3.0 \pm  0.4$ &  $ 0.97^{+ 0.42}_{- 0.31}$ &  1.2 (21) &  3.3 $\pm$  0.8 \\
 & 31174017 &  4.3 & $ 2.7 \pm  1.0$ &  $ 0.96^{+ 1.05}_{- 0.42}$ &  0.2 (5) &  2.9 $\pm$  1.2 \\
 & 31174003 &  4.3 & $ 3.0 \pm  0.5$ &  $ 1.22^{+ 0.67}_{- 0.40}$ &  0.6 (9) &  2.2 $\pm$  0.5 \\
 & 31174010 &  4.3 & $ 2.9 \pm  0.4$ &  $ 1.07^{+ 0.49}_{- 0.33}$ &  0.6 (18) &  3.4 $\pm$  0.8 \\
\hline
4 & 31174018 &  4.3 & $ 2.4 \pm  0.4$ &  $ 1.02^{+ 0.31}_{- 0.23}$ &  0.5 (10) &  4.6 $\pm$  1.2 \\
 & 31174009 &  4.3 & $ 2.5 \pm  0.3$ &  $ 1.15^{+ 0.31}_{- 0.24}$ &  1.0 (15) &  4.1 $\pm$  0.8 \\
 & 31174014 &  4.3 & $ 2.3 \pm  0.4$ &  $ 0.96^{+ 0.27}_{- 0.20}$ &  0.7 (11) &  3.7 $\pm$  0.8 \\
 & 31174007 &  4.3 & $ 2.5 \pm  0.3$ &  $ 1.43^{+ 0.35}_{- 0.27}$ &  0.9 (12) &  3.6 $\pm$  0.6 \\
 & 31174013 &  4.3 & $ 2.4 \pm  0.3$ &  $ 1.50^{+ 0.34}_{- 0.27}$ &  1.2 (12) &  4.1 $\pm$  0.7 \\
 & 31174012 &  4.3 & $ 2.3 \pm  0.2$ &  $ 1.42^{+ 0.24}_{- 0.20}$ &  1.5 (28) &  5.5 $\pm$  0.9 \\
 & 31174008 &  4.3 & $ 2.2 \pm  0.3$ &  $ 1.45^{+ 0.30}_{- 0.24}$ &  0.4 (18) &  4.1 $\pm$  0.7 \\
 & 31174011 &  4.3 & $ 2.0 \pm  0.2$ &  $ 1.42^{+ 0.20}_{- 0.18}$ &  0.9 (16) &  6.3 $\pm$  1.2 \\
 & 31174016 &  4.3 & $ 2.0 \pm  0.2$ &  $ 1.55^{+ 0.20}_{- 0.18}$ &  1.0 (14) &  5.4 $\pm$  0.8 \\
\hline
5 & 31174021 &  4.3 & $ 1.5 \pm  0.1$ &  $ 10.20^{+ 0.42}_{- 0.42}$ &  1.0 (80) &  7.5 $\pm$  0.5 \\
 & 31174019 &  4.3 & $ 1.5 \pm  0.1$ &  $ 15.79^{+ 0.71}_{- 0.71}$ &  1.2 (67) &  8.5 $\pm$  0.7 \\
 & 31174020 &  4.3 & $ 1.6 \pm  0.1$ &  $ 18.68^{+ 0.99}_{- 0.97}$ &  0.6 (53) &  6.6 $\pm$  0.5 \\
\hline
6 & 31174002$^{\textrm{PC}}$ &  4.3 & $ 1.2 \pm  0.1$ &  $ 33.49^{+ 1.79}_{- 1.78}$ &  0.8 (47) &  7.5 $\pm$  0.7 \\
 & 31174002$^{\textrm{WT}}$ &  4.3 & $ 1.3 \pm  0.3$ &  $ 37.51^{+ 3.60}_{- 3.59}$ &  0.9 (18) &  6.1 $\pm$  1.1 \\
 & 31174001$^{\textrm{PC}}$ &  4.3 & $ 1.3 \pm  0.1$ &  $ 48.33^{+ 2.91}_{- 2.91}$ &  1.2 (36) &  8.1 $\pm$  0.8 \\
 & 31174001$^{\textrm{WT}}$ &  4.3 & $ 1.4 \pm  0.1$ &  $ 51.33^{+ 2.14}_{- 2.13}$ &  1.0 (94) &  6.0 $\pm$  0.5 \\
\hline
7 & 31166001 & $ 4.2^{+ 0.4}_{- 0.4}$ & $ 1.6 \pm  0.1$ &  $253.13^{+ 19.75}_{- 16.00}$ &  0.9 (139) &  6.2 $\pm$  0.3 \\
\footnotetext[1]{Unabsorbed 0.5-10 keV luminosity in units of $10^{35}$ erg s$^{-1}$}
\footnotetext[2]{C-statistic}
\end{tabular}
\end{table}

\subsection{Model 2: Soft emission and thermal contributions}
\label{sec:thermal}

\begin{figure*}
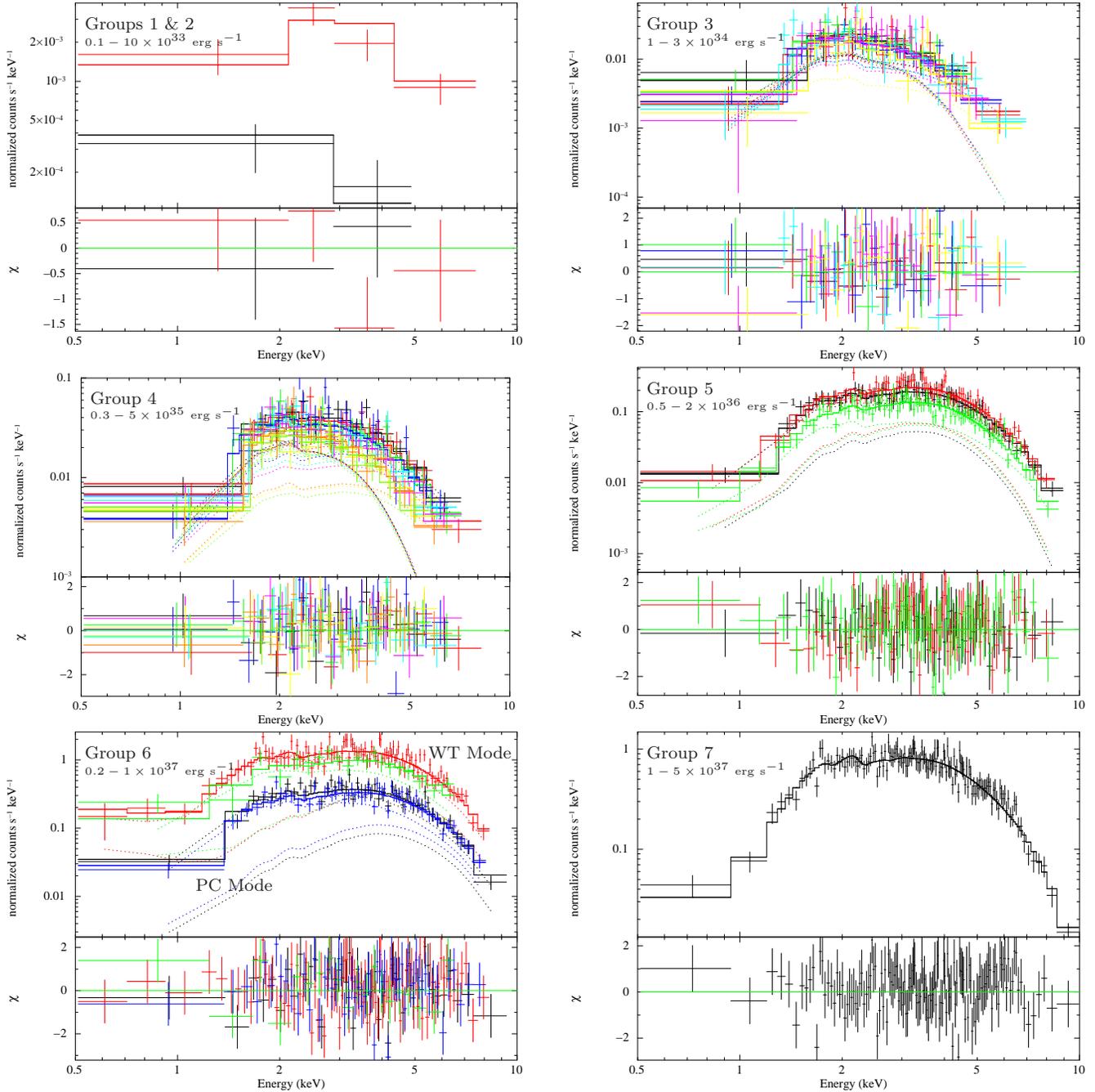

\begin{tabular}{c c}
\begin{overpic}[width=0.325\textwidth,angle = 270]
{bin12_pl_nh43_tbabs_newfit.ps}
  \put(15, 63){Groups 1 \& 2} \put (15,60) {\tiny $0.1-10\times10^{33}$ erg s$^{-1}$} \end{overpic} &
\begin{overpic}[width=0.325\textwidth,angle = 270]
{bin3_plbb_nh43_tbabs_add.ps}
 \put (15,63) {Group 3} \put (15,60) {\tiny $1-3\times10^{34}$ erg s$^{-1}$}\end{overpic} \\
\begin{overpic}[width=0.325\textwidth,angle = 270]
{bin4_plbb_nh43_tbabs_add.ps}
 \put (15,63) {Group 4} \put (15,60) {\tiny$0.3-5\times10^{35}$ erg s$^{-1}$}\end{overpic} &
\begin{overpic}[width=0.325\textwidth,angle = 270]
{bin5_plbb_nh43_tbabs_add.ps}
 \put (15,63) {Group  5} \put (15,60) {\tiny$0.5-2\times10^{36}$ erg s$^{-1}$}\end{overpic} \\
\begin{overpic}[width=0.325\textwidth,angle = 270]
{bin6_plbb_nh43_tbabs_add.ps}
 \put (15,63) {Group 6} \put (15,60) {\tiny$0.2-1\times10^{37}$ erg s$^{-1}$}
 \put (80,63) {WT Mode} \put(36, 38) {PC Mode} \end{overpic} &
\begin{overpic}[width=0.325\textwidth,angle = 270]
{bin7_pl_nh43_tbabs.ps}
 \put (15,63) {Group 7} \put (15,60) {\tiny$1-5\times10^{37}$ erg s$^{-1}$}\end{overpic} \\
\end{tabular}
\caption{\emph{Swift} XRT folded spectra of SAX J1750.8-2900 used in this work, together with our best-fit model functions. Observations were combined into 7 groups (Sec. \ref{sec:specanalysis}, Table \ref{tab:groupfits}). Groups 1, 2 and 7 are fit with an absorbed power law, while with groups 3-6 we were able to constrain an additional soft component, and so we plot their \texttt{powerlaw + bbodyrad} fits. Adding a soft component improved residuals slightly, but overall the spectra and their residuals are similar between the single and multicomponent fits. Groups 1 and 2 are plotted together but are fit separately and binned for display only. \protect{ \label{fig:binspectra}} }
\end{figure*}

We were able to fit a soft component (modeled with a blackbody) in addition to the hard emission (modeled with a powerlaw) when fitting several of the groups of observations. There was no change in the chi-squared value compared to the pure power law fit, indicating there is no strong model preference. The multicomponent fit parameters are shown in Table \ref{tab:groupfits} and spectra with their group fits (either powerlaw or powerlaw plus blackbody models) are shown in Figure \ref{fig:binspectra}.

%BINNED SPECTRAL PLOTS

Despite group 7 being the brightest observation, it had fewer source counts than any of groups 3, 5 or 6. We were unable to calculate the lower error bound on the blackbody temperature when we performed a powerlaw plus blackbody fit, indicating the blackbody component was not definitively detected in this observation. For a blackbody normalization of 5 (an average of the normalizations found in fits on groups 3-6), the blackbody had a temperature upper limit of 2.45 keV and maximal flux contribution of 25\%.

\begin{figure}
\begin{center}
\begin{overpic}[width=0.5\textwidth]
{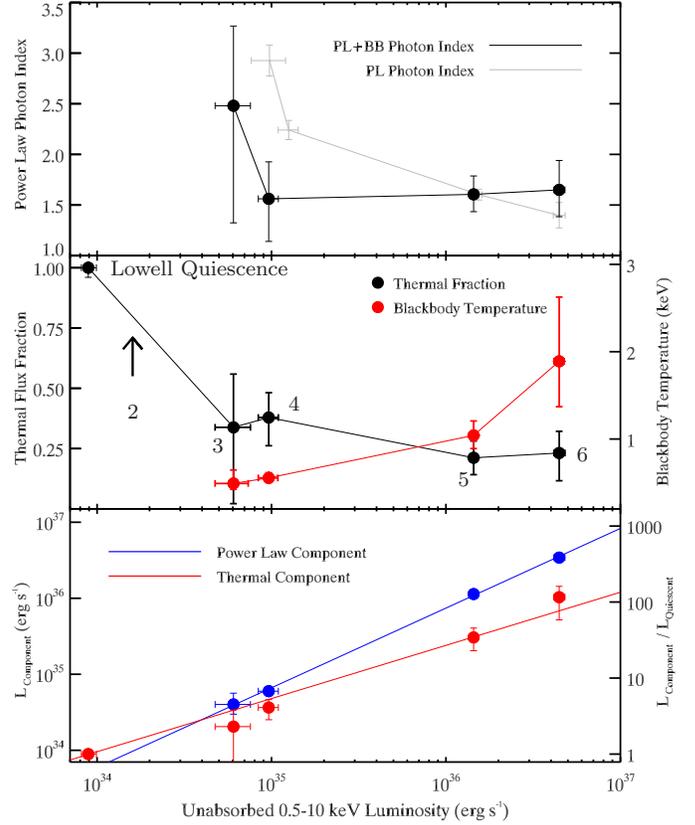}
 \put (12,65) {Lowell Quiescence} \put (14,48) {2} \put (24,44) {3}  \put (33,49) {4} \put (53,40) {5} \put (67,43) {6}   \end{overpic}
\caption{\textbf{Model 2, Power law + Blackbody}: The top panel shows the difference in the power law photon index behavior between the two models: pure power law (plotted in grey) and power law plus blackbody (black). When a soft component (i.e. blackbody) is included in the fit, the photon index is essentially constant between $L_X = 4\times10^{36} \rightarrow 10^{35}$ erg s$^{-1}$. In the middle panel are the thermal fraction (black) and blackbody temperature (red) as a function of the total 0.5-10 keV unabsorbed luminosity. The soft component's temperature is lower at lower luminosities, which is expected if the soft component is associated with accretion onto the neutron star surface. Bottom panel: Luminosity from each of the power law (blue) and blackbody (red) components in the power law plus blackbody model plotted along with a linear fit, along with the \citet{lowell} \texttt{NSATMOS} thermal luminosity (at $\sim10^{34}$ erg s$^{-1}$). The power law flux declines more rapidly than the flux from the thermal component. \protect{ \label{fig:pplbbconstraints}} }
\end{center}
\vspace{-4mm}
\end{figure}

Luminosity-photon index relations for the power law and the power law plus blackbody models are compared in the top panel of Figure \ref{fig:pplbbconstraints}. With the addition of the soft component, between $4\times 10^{36}$ and $1 \times 10^{35}$ erg s$^{-1}$ the photon index was consistent with being constant ($\Gamma \sim 1.6$). At a luminosity of $L_X = 6 \times 10^{34}$  erg s$^{-1}$, however, the power law component softened ($\Gamma = 2.5^{+0.8}_{-1.2}$) although the error bars are large and still consistent  with the photon index at higher luminosities.

Shown in the middle panel of Figure \ref{fig:pplbbconstraints}, the blackbody component exhibited a decreasing temperature ($1.9 \rightarrow 0.5$ keV, plotted in red) at lower luminosities ($L_X \sim 4 \times10^{36} \rightarrow 6 \times 10^{34}$ erg s$^{-1}$). The blackbody's contribution to the total flux (i.e. the thermal flux fraction, plotted in black) tends towards higher fractions at lower luminosities, although the uncertainties are large. We could only fit for an additional soft component in groups 3-6, but we have also plotted the quiescent thermal fraction (with a lower limit of 96\% as the quiescent spectrum was thermal with a maximum power law contribution of 4\%) and luminosity from the \citet{lowell} results. We estimated the lower limit to the thermal fraction of the flare by simply dividing the quiescent thermal luminosity by the flare luminosity.

In the bottom panel of Figure \ref{fig:pplbbconstraints}, we plot the power law and blackbody components' 0.5-10 keV luminosities as a function of the total 0.5-10 keV luminosity. When we performed a simple linear fit, we found the power law flux declines more rapidly than the thermal flux (a slope of 1.1 versus 0.70 for the blackbody flux), which is consistent with the thermal flux fraction increasing towards lower luminosities. If we express each component's luminosity in terms of the quiescent thermal luminosity (right axis, bottom panel), we find that for groups 3 and 4 the thermal luminosity is 2-4 times the quiescent luminosity, indicating that the quiescent emission contributes significantly to the thermal luminosity (approximately 50\% and 25\% of the thermal luminosity at $6\times 10^{34}$ erg s$^{-1}$ and $1 \times 10^{35}$ erg s$^{-1}$, respectively).

%PL + BB Constraints

\section{Discussion}
Our set of \emph{Swift} XRT observations of SAX J1750.8-2900 span over 4 years and 3 orders of magnitude in luminosity, which allows us to explore a wide range of accretion rates and timescales. We find that the source softens between outburst and quiescence, $L_X = 10^{-2} \rightarrow 10^{-3}$ L$_{NS}^{Edd}$, and the softening mechanism generates spectral variability on multi-day timescales.

\subsection{Spectral softening between outburst and quiescence}
\label{sec:discussoft}

Spectral softening is a nearly universal behavior in black hole transient systems as they enter quiescence \citep{wugubh,plotkin} and now a trend seen across dozens of neutron star transients \citep{wijnandssoft}. In black hole binaries, one interpretation of the spectral softening during the transition from the hard state to quiescence involves the changing opacity of an inner advection-dominated accretion flow (ADAF), although jets may also play a role in the softening near quiescence. The anti-correlation between the photon index and luminosity typically spans $L_X /  L_{BH}^{Edd}$  = $10^{-2} \rightarrow 10^{-5}$ for $\sim 10 \ M_{\odot}$ black holes, while below $L_X /  L_{BH}^{Edd}$ = $10^{-5}$, there is evidence the spectrum does not soften further into quiescence but that the photon index remains constant \citep{plotkin}.

While both neutron star and black hole transients exhibit increasing photon indices towards lower luminosities, the softening mechanism(s) are not necessarily expected to be the same, largely due to the differences in emission sources between the two types of transients. Gathering observations of neutron star transients with luminosities between $10^{34}-10^{36}$ erg s$^{-1}$ and comparing to the photon index-luminosity relation for black holes found by \citet{plotkin}, \citet{wijnandssoft} found that softening black hole and neutron star transients are distinctly separate populations. Black holes are significantly harder ($\Gamma \lesssim 2$) than neutron star transients ($2 \lesssim \Gamma \lesssim 3$) below $10^{35}$ erg s$^{-1}$.

With the exception of accretion disk emission, black hole transients are almost entirely sources of nonthermal emission and can be modeled with a pure power law at luminosities below an Eddington fraction of 1\%. Neutron star LMXBs, however, have both nonthermal and thermal emission, although it is unclear which component dominates in different luminosity and accretion regimes. Due to the neutron star's solid surface, there are multiple sources of soft X-ray emission across a range of luminosities: boundary layer emission \citep{linbl}, thermal surface emission due to heat generated by pycnonuclear reactions deep within the neutron star crust \citep{brownpycnonuclear}, and a blackbody-like spectrum due to low-level accretion \citep{zampieri}, while accretion can produce nonthermal emission at almost any luminosity. Changes in multiple or just one of these sources can contribute to the spectral softening. Attempts to track the thermal and nonthermal contributions as a function of time and luminosity are further complicated when there is insufficient spectral quality to constrain both components. Despite these obstacles, there is growing evidence suggesting that changes in a soft component's properties is largely responsible for spectral softening \citep{igrj17494,terzan5,wijnandssoft} and that the component is powered by accretion rather than heat release from the neutron star crust, while there may be simultaneous changes in the nonthermal emission associated with accretion.

\begin{figure}
\begin{center}
\includegraphics[scale = 0.75]{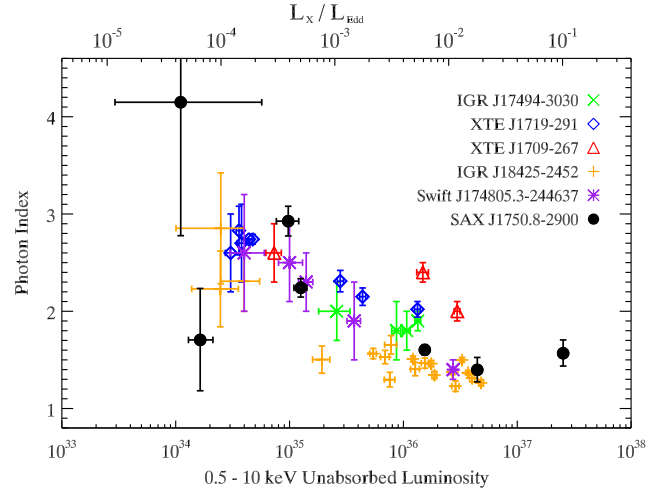}
\caption{Power law photon indices versus the unabsorbed 0.5-10 keV luminosity obtained from pure power law fits for multiple neutron star LMXBs that exhibit spectral softening with decreasing luminosity, or hardening with increasing luminosity in the case of Swift J174805.3-244637 \citep{igrj17494,padillaxtej1719,dgenaarxte1709,m28soft,terzan5}. \protect{\label{fig:softsources}} }
\end{center}
\vspace{-4mm}
\end{figure}

Softening has been observed in neutron star LMXBs over luminosities of $10^{36} \rightarrow 10^{34}$ erg s$^{-1}$, similar to the range where we have detected softening in SAX J1750.8-2900 between outburst and quiescence. In Figure \ref{fig:softsources} we have plotted the photon indices and luminosities for SAX J1750.8-2900's power law fits, along with the published power law parameters for several other NS LMXBs that have exhibited spectral softening.  For a more complete comparison of spectral softening in a dozen neutron star transients, see Figure 1 in \citet{wijnandssoft}.

\subsubsection{Soft Component's Role in Softening}

In most of the neutron star softening sources, a power law provided adequate, if not good, fits to most of the observations. However, at lower luminosities ($\lesssim 10^{35}$ erg s$^{-1}$) and with higher SNR spectra, many sources required an additional soft spectral component, indicating the spectra are not entirely non-thermal, and the softening cannot be accurately modeled by a power law with a steepening photon index, as is the case with softening black hole transients.

In deep \emph{XMM} observations of neutron star transients XTE J1709-267 and IGR J1794-3030, \citet{dgenaarxte1709} and \citet{igrj17494} achieved the sensitivity to constrain both hard and soft spectral components; both sources exhibited a decaying thermal temperature and constant photon index while the total flux decreased over the course of approximately 10 hours. Due to the similarities between the two systems' behaviors, \citet{igrj17494} credited the spectral changes in both sources to a decrease in the mass accretion rate onto the neutron star surface which produced a soft component with a decreasing temperature, while the nonthermal flux due to accretion also decreases. 

With SAX J1750.8-2900 when adding a blackbody, we found almost no change in the chi-squared values, indicating there was no model preference between a pure power law and the power law plus blackbody and that the additional soft component was not statistically required. Thus, the softening in SAX J1750.8-2900 from our data can be accurately described by a steepening photon index towards lower luminosities. However, it is worth noting that when we included a blackbody component, we find the soft component's behavior is similar to that seen in other softening neutron star transients where additional soft components are statistically required, and the behavior is consistent with the interpretation that the soft component is generated by accretion onto the neutron star. In SAX J1750.8-2900, as the total luminosity decreased, the blackbody displayed lower temperatures and its flux contribution increased, indicating the strength of the component was increasing relative to the hard component. Between $L_X \sim 10^{36}$ and $10^{35}$ erg s$^{-1}$, the photon index of the power law component was essentially constant revealing that an increasingly dominant thermal component may contribute significantly to the softening. We are unable to comment on the behavior of the power law component's photon index above $L_X \sim5 \times 10^{36}$ erg s$^{-1}$ and below $\sim5\times10^{34}$ erg s$^{-1}$, as we either were unable to constrain both power law and blackbody components or the photon index suffered from large uncertainties. 

%WHERE TO ADD COMMENT ABOUT COMMON ORIGIN ABOVE X LUM DUE TO LACK OF CHANGE IN PARAM

%Binned Power Law Fit Parameters - Plot

\subsubsection{Variability and Spectral Hardening}

Rapid, non-monotonic variability, as well as alternating spectral softening and hardening with decreasing and increasing luminosity seen in SAX J1750.8-2900 strongly support the theory that accretion generates the spectral softening. With our dense set of observations between August and October 2008 (Figure \ref{fig:posinglefitplot2008}), we found SAX J1750.8-2900's luminosity and spectral hardness varied on timescales as short as a day; such rapid variability can be generated by changes in the accretion rate onto the neutron star. Additionally, SAX J1750.8-2900 displays alternating spectral hardening and softening over the course of days and weeks, which has not been seen in other neutron stars on such timescales.

Spectral hardening with increasing intensity has been observed in the neutron star LMXB Swift J174805.3-244637 as its rise to the hard state was monitored with \emph{Swift} XRT \citep{terzan5}. As its luminosity increased from $L_X = 4\times10^{34} \rightarrow 10^{36}$ erg s$^{-1}$ ($\sim 2\times10^{-4} \rightarrow 4 \times 10^{-3}$ L$_{Edd}$) the photon index decreased ($\Gamma = 2.6 \rightarrow 1.7$) when fit with a pure power law. Using more complex models with a soft component, \citet{terzan5} found an increasing thermal temperature with increasing luminosity, which cannot be produced via a cooling neutron star crust since the observations are at the onset outburst. While we could not track the rapid day-to-day variations in the soft component as SAX J1750.8-2900 hardened, the similarities between SAX J1750.8-2900 and Swift J174805.3-244637 further support the interpretation that the spectral changes are produced by hard and soft emission powered by a varying mass accretion rate onto the neutron star surface.

While we found SAX J1750.8-2900 to be variable during August to October 2008, its average luminosity was approximately $10^{35}$ erg s$^{-1}$, which shows that underluminous accretion flows ($L_X / L_{\textrm{Edd}} <0.1 \%$) can be stable under certain conditions. Namely, SAX J1750.8-2900 remained at $L_X \sim10^{35}$ erg s$^{-1}$ for over one month (Figure \ref{fig:posinglefitplot2008}) in contrast with the rapid decay to quiescence ($\lesssim$ 1 week) seen in other neutron star transients (Aql X-1, \citet{aqlx1campana}; SAX J1810.8-2069 \citet{atel1260}; 4U 1608-52 \citet{atel2264}).

\subsection{Accretion and the Origin of the Soft Component}
\label{sec:thermalorigin}

In many of the existing discussions of softening neutron star LMXBs, emission associated with accretion onto the neutron star surface has been cited as the possible soft component detected in the spectra. The spectrum generated by spherical accretion onto an unmagnetized neutron star has been a long-standing issue. \citet{zeldovich} demonstrated that freely, radially infalling ions onto a neutron star produced a blackbody spectrum with a high-energy tail due to Comptonization with luminosities of $\sim10^{35}-10^{38}$ erg s$^{-1}$. Similar results were obtained by subsequent studies even when additional physics were included. \citet{zampieri} extended simulations to even lower mass accretion rates down to luminosities of $\sim10^{31}-10^{32}$ erg s$^{-1}$ and found the emergent spectra were significantly hardened compared to a blackbody at the neutron star effective temperature. \citet{deufel} considered the case of non-spherical accretion from a hot accretion flow (ADAF) surrounding the neutron star. The hot ions heat the surface layer to produce a blackbody-like spectra with a high-energy Comptonized tail, highlighting the contribution of neutron star surface to the hard emission in LMXBs. Further investigations are required to determine how the surface emission is modified by the accretion flow. But in all cases, accretion onto the neutron star surface produces a hardened blackbody-like spectrum and, at high accretion accretion rates, a significant high-energy tail.

We note that at luminosities above $L_X \sim 10^{36}$ erg s$^{-1}$, corresponding to groups 5-7, there is not a significant change in the hardness ratio or spectral parameters (photon indices, blackbody temperatures, etc.) for either the powerlaw or powerlaw plus blackbody models, despite over an order of magnitude change in luminosity. This 'plateau' in spectral parameters could suggest that the emission, whether entirely nonthermal or a combination of thermal and nonthermal sources, above $10^{36}$ erg s$^{-1}$ only changes in brightness without a substantial change in the accretion flow properties.

\subsection{Sub-Outburst Accretion and Crustal Heating Implications}

Accretion in the outburst state dominates the thermal evolution of transiently accreting neutron stars. During outburst, the accreted material compresses the upper layers of the neutron star, inducing pycnonuclear reactions in a process known as deep crustal heating \citep{brownpycnonuclear}. The heat generated by the reactions is conducted throughout the neutron star and is partially radiated away at the surface, emerging as a thermal spectrum. From the time-averaged outburst accretion rate, which is determined by the outburst history, one can estimate the neutron star's quiescent luminosity and surface temperature due to deep crustal heating. Time-averaged accretion rates due to sub-outburst accretion, however, are less well-known as there are typically large uncertainties in the accretion events' properties, such as duration, intensity and recurrence times as this type of low-level activity is difficult to observe with all-sky X-ray monitors.

The heating problem is particularly important in SAX J1750.8-2900 because it exhibits two forms of accretion outside of outburst: sustained ($>$ 1 month) accretion at $10^{35}$ erg s$^{-1}$ and a low-level flare in quiescence, while also having one of the highest reported surface temperatures for a neutron star in quiescence \citep{lowell}. The system warrants an investigation as to whether these forms of low-level accretion could be frequent enough to produce shallow heating in the crust and contribute to the high thermal luminosity.

The intensity and type of accretion flow that manifests in quiescent transients has been a long-standing issue. A hard power law component often seen in quiescent neutron star spectra has commonly been attributed to ongoing accretion. It has been recently proposed that the hard, high-energy tail is produced by thermal bremsstrahlung emission from a radiatively-inefficient accretion flow \citep{cenx4}. SAX J1750.8-2900, however, does not show signs of nonthermal emission in quiescence nor does it show signs of spectral variability, which is another sign of ongoing accretion. The 2010 and 2012 quiescent luminosities were consistent \citep{lowell, saxjflare} further supporting the claim that accretion is not continuous during quiescence in SAX J1750.8-2900, although the \citet{saxjflare} luminosity has large uncertainties.

While SAX J1750.8-2900 does not exhibit the characterisitic signs of continuous accretion in quiescence, the detection of a flare \citep{saxjflare} suggests accretion is at least sporadic. Flaring activity of similar intensity and duration as in SAX J1750.8-2900 has been exhibited in several X-ray neutron star transient systems, including KS 1741-293 \citep{degenaarks1741} and XTE 1701-462 \citep{fridriksson_2011}. While we lack the statistics to monitor the progression of the flare in SAX J1750.8-2900, \citet{fridriksson_2011} studied the hardness-intensity diagram of a flare observed in XTE 1701-462 that  reached a peak luminosity 20 times higher than the system's quiescent level. Both the thermal and non-thermal fluxes increased, albeit differently, strongly suggesting the flare was accretion-powered. 

With our \emph{Swift} XRT constraints on the accretion events (the quiescent flare and $L_X =10^{34}-10^{36}$ erg s$^{-1}$ activity) we can estimate their associated time-averaged accretion rates and compare to that of outburst, which is expected dominate the thermal evolution of the neutron star. We roughly approximated the flare as having a duration of 16 days (the longest possible duration in order to calculate an accretion rate upper limit) with a constant flux ($\sim4 \times 10^{-12}$ erg s$^{-1}$ cm$^{-2}$- our 0.5-10 keV unabsorbed flux estimate for the flare) and a (highly unconstrained) recurrence rate of 3 times per year. Bolometric corrections are model and luminosity dependent, but we use the ratio between the unabsorbed bolometric and 0.5-10 keV quiescent fluxes reported by \citet{lowell} as a correction estimate (B.C. = 1.2 for the \texttt{NSATMOS} fluxes). After we convert the flare's flux to a bolometric luminosity, we subtract the bolometric luminosity associated with the neutron star surface emission in quiescence (i.e. the \citet{lowell} quiescent luminosity). Assuming an accretion efficiency of $\epsilon = 0.2$, we find a time-averaged accretion rate of $\dot{M} = 1\times10^{-13} \ M_{\odot}$ per year (using $L_{Bol} = \epsilon \dot{M} c^2$); even a recurrence rate of 10 flares per year only yields $\dot{M} = 3\times 10^{-13}$ M$_{\odot}$ per year which is nearly 2 orders of magnitude smaller than our outburst averaged accretion rates, $\dot{M} = 0.4-2.2 \times 10^{-10}$ M$_{\odot}$/year.

We estimated the outburst $\dot{M}$ based on an outburst duration of 4 months with an occurrence rate of once every 4 years. We have detected SAX J1750.8-2900 in outburst with luminosities between $L_X = 0.4 - 2.5 \times 10^{37}$ erg s$^{-1}$, corresponding to range of accretion rates previously stated. \citet{lowell} computed an outburst mass accretion rate of $2 \times 10^{-10} \ M_{\odot}$ per year which is in agreement with our estimates. For SAX J1750.8-2900's activity between August and October 2008, we estimated a maximum rate of $\dot{M} = 2 \times 10^{-12} \ M_{\odot}$ per year based on a duration of 60 days (August and September) with a typical $L_X = 10^{35}$ erg s$^{-1}$ and a recurrence time of once per year. If this behavior is only associated with outbursts, then we'd expect a lower occurrence rate (once per 4 years) and a lower mass accretion rate ($\sim 4\times 10^{-13} \ M_{\odot}$).

Although there are large uncertainties in our assumptions (recurrence times, durations, etc.) we find that sub-outburst accretion events lead to mass accretion rates \emph{at least} an order of magnitude lower than rates associated with the outburst state. This supports the claim that the thermal evolution of the neutron star in SAX J1750.8-2900 is most heavily dependent on outburst accretion and that low-level accretion has little to no effect on the long term crustal temperature. Our conclusion is in agreement with \citet{fridriksson_2011}, who performed a similar analysis based on flares exhibited by XTE J1701-462 and found that low-level accretion is unlikely to have a significant effect on the equilibrium surface temperature. We do not, however, address the immediate crustal heating effects due to small accretion events, such as flares, as they require much more detailed observations and calculations.

\section{Conclusions}

In our \emph{Swift} XRT study of SAX J1750.8-2900's X-ray spectral behavior over four years of activity, we have found that the source softens as its luminosity decreases between $L_X=4\times10^{36} \rightarrow 10^{35}$ erg s$^{-1}$. This trend is consistent with the spectral softening observed in individual neutron star LMXB systems, as well as the cumulative softening behavior exhibited by over one dozen transients between the luminosities of $10^{36}$ and $10^{34}$ erg s$^{-1}$. Our data is consistent with both the softening being due to a steepening power law towards lower luminosities and a thermal component becoming more apparent in the spectrum. The soft component may be associated with accretion onto the neutron star surface and has been definitively detected in other softening neutron star transients, which supports the interpretation that the thermal emission is driving the softening as the luminosity decreases.

\section{Acknowledgements}
We thank Mike Nowak for his useful comments and discussions. JA and ML acknowledge funding from a NASA-Swift award (NNX12AE60G, PI: Linares).

\newpage

\bibliography{references}

\end{document}